\documentclass[aps,preprintnumbers,floatfix,article,amsmath,amssymb,floatfix,10pt,prd,superscriptaddress,nofootinbib]{revtex4}
\usepackage{bm}
\usepackage{amsfonts}
\usepackage{latexsym}
\usepackage{graphicx}
\usepackage{amsmath}
\usepackage{palatino}
\usepackage{mathpazo}
\usepackage{textcomp}
\usepackage{float}
\usepackage{booktabs}
\usepackage{dcolumn}
\usepackage{ragged2e}
\usepackage{hyperref}
\hypersetup{colorlinks,citecolor=magenta}
\hypersetup{colorlinks=true,linkcolor=red,filecolor=magenta,urlcolor=blue}
\usepackage{amsmath}
\usepackage{xcolor}
\usepackage{epsfig}
\usepackage{caption}
\usepackage{subcaption}
\usepackage{commath}
\def\jnl@style{\it}
\def\aaref@jnl#1{{\jnl@style#1}}

\def\aaref@jnl#1{{\jnl@style#1}}

\def\aj{\aaref@jnl{AJ}}                   
\def\apj{\aaref@jnl{ApJ}}                 
\def\apjl{\aaref@jnl{ApJ}}                
\def\apjs{\aaref@jnl{ApJS}}               
\def\apss{\aaref@jnl{Ap\&SS}}             
\def\aap{\aaref@jnl{A\&A}}                
\def\aapr{\aaref@jnl{A\&A~Rev.}}          
\def\aaps{\aaref@jnl{A\&AS}}              
\def\mnras{\aaref@jnl{Mon.~Not.~Roy.~Astron.~Soc.}}             
\def\prd{\aaref@jnl{Phys.~Rev.~D}}        
\def\prc{\aaref@jnl{Phys.~Rev.~C}}  
\def\prl{\aaref@jnl{Phys.~Rev.~Lett.}}    
\def\qjras{\aaref@jnl{QJRAS}}             
\def\skytel{\aaref@jnl{S\&T}}             
\def\ssr{\aaref@jnl{Space~Sci.~Rev.}}     
\def\zap{\aaref@jnl{ZAp}}                 
\def\nat{\aaref@jnl{Nature}}              
\def\aplett{\aaref@jnl{Astrophys.~Lett.}} 
\def\apspr{\aaref@jnl{Astrophys.~Space~Phys.~Res.}} 
\def\physrep{\aaref@jnl{Phys.~Rep.}}      
\def\physscr{\aaref@jnl{Phys.~Scr}}       
\def\commat{\aaref@jnl{Comm.~Math.~Phys.}}              
\def\science{\aaref@jnl{Science}}               
\def\cqg{\aaref@jnl{Classical Quant.~Grav.}}            
\def\jpcs{\aaref@jnl{JPCS}}                                     
\def\ijmpd{\aaref@jnl{Int.~J.~Mod.~Phys.~D}}                    
\def\grg{\aaref@jnl{Gen.~Relat.~Gravit.}}               
\def\rpp{\aaref@jnl{Rep.~Prog.~Phys.}}          
\def\npa{\aaref@jnl{Nucl.~Phys.~A}}        
\def\lrr{\aaref@jnl{Living Rev.~Rel.}}                   
\def\jcap{\aaref@jnl{J.~Cosmology Astropart.~Phys.}}    
\def\rmp{\aaref@jnl{Rev.~Mod.~Phys.}}   
\def\epjc{\aaref@jnl{Eur.~Phys.~J.~C}} 
\def\plb{\aaref@jnl{~Phy.~Lett.~B}} 
\def\mpla{\aaref@jnl{Mod.~Phy.~Lett.~A}} 
\def\arxiv{\aaref@jnl{arxiv.org}}

\allowdisplaybreaks[1]

\begin{document}
\color{black}       
\title{\bf Dynamical system analysis for accelerating models in non-metricity $f(Q)$ gravity}

\author{S. A. Narawade}
\email{shubhamn2616@gmail.com}
\affiliation{Department of Mathematics,
Birla Institute of Technology and Science-Pilani, Hyderabad Campus,
Hyderabad-500078, India.}

\author{Laxmipriya Pati}
\email{lpriyapati1995@gmail.com}
\affiliation{Department of Mathematics,
Birla Institute of Technology and Science-Pilani, Hyderabad Campus,
Hyderabad-500078, India.}

\author{B. Mishra}
\email{bivu@hyderabad.bits-pilani.ac.in }
\affiliation{Department of Mathematics,
Birla Institute of Technology and Science-Pilani, Hyderabad Campus,
Hyderabad-500078, India.}

\author{S.K. Tripathy}
\email{tripathy\_sunil@rediffmail.com }
\affiliation{Department of Physics, Indira Gandhi Institute of Technology, Sarang, Dhenkanal, Odisha-759146, India.}


\begin{abstract}

Two accelerating cosmological models are presented in symmetric teleparallel $f(Q)$ gravity, $Q$ be the non-metricity. The models are constructed based on the assumptions of two different functional forms of $f(Q)$ and a dynamically changing nature of the deceleration parameter that shows transition at $t=2n\pm\sqrt{\frac{4n^2+1}{3}}$, $n$ being a positive constant. In both the models, the equation of state parameter for the dark energy in $f(Q)$ gravity becomes a dynamical quantity and crosses the phantom divide line. The violation of the strong energy condition and the null energy condition at late times are also established. In addition, the dynamical system analysis has been performed and three critical points in each model are identified. In each model, at least one stable node has been observed. To strengthen further, the stability analysis using homogeneous linear perturbations has been performed to ensure the stability of the models.
\end{abstract}

\maketitle

\section{Introduction} 

The geometrical modification of General Relativity (GR) has become inevitable post supernovae observations and other cosmological observations \cite{Riess98,Perlmutter99,Ade16,Aghanim16}. GR has been described in terms of the Levi-Civita connection by complying the basis of Riemannian geometry. In this framework, the Ricci curvature $R$ has been considered as the building block of the space-time. In this approach, the geometry is free from torsion and non-metricity. Apart from Riemannian geometry, GR can also be described in terms of other geometries, one among them is the teleparallel gravity \cite{Aldrovandi14}. In this approach, the gravitational force is driven by the torsion $\mathcal{T}$ instead of curvature $R$ as in Riemannian geometry. Einstein had used this geometry in his attempt of unified field theory, but formally this geometry has been introduced in \cite{Einstein28}. Another approach would be the non-metricity approach, where the non-metricity $Q$ mediates the gravitational interaction and is free from curvature and torsion \cite{Nester99}, known as the symmetric teleparallel gravity. To mention here, rather than the physical manifold, the mechanism mediating gravity is the affine connection. In GR, the curvature is used, which is a property of connection, but not of the manifold. In the same approach the non-metricity is another connection property. The non-metricity $Q$ turns out to be the Einstein pseudo-tensor, which becomes a true tensor in the geometric representation. Further Jimenez et al. \cite{Jimenez18} have developed the symmetric teleparallel gravity into coincident GR or $f(Q)$ gravity. Since in modern cosmology the interest is on the extended theories of gravity, therefore the alternative geometries are being explored. This is being done with the belief that the features of gravitational theories would be different corresponding to the  Riemannian geometry and get some insight into the late time cosmic acceleration issue \cite{Ferraro08,Geng11,Cai16,Jarv16}. Another interesting note is to generalize the metric theories to a more general geometry from Riemannian geometry as can be seen in ref. \cite{Conroy18}. To note here Harko et al. \cite{Harko18} have given the modified $f(Q)$ gravity by considering an extension of symmetric teleparallel gravity in the framework of metric-affine formalism. \\

In the recent literature on extended theories of gravity, the cosmological and astrophysical aspects of symmetric teleparallel gravity or $f(Q)$ gravity is being given importance. The propagation velocity of gravitational waves around Minkowski space-time and their potential polarizations was investigated in the general class of symmetric teleparallel gravity \cite{Hohmann19}. In this gravitational theory, the role of gravitational wave polarizations has been significant in restricting the gravitational theories strong field behaviour \cite{Soudi19}. In $f(Q)$ gravity, Lazkoz et al. \cite{Lazkoz19} have reformulated the  Lagrangian of $f(Q)$ as an explicit function of redshift and performed the observational analysis of $f(Q)$ models with the cosmological data sets. It also gives the explanation to the cosmic expansion and can explain that the dark energy can be identified by the space-time. The equation of state of the geometrical dark energy can cross over the phantom divide line \cite{Lazkoz19}. With the order reduction method, the functional form of $f(Q)$ could be constrained and by integrating the field equations, the bouncing cosmology model can be realised \cite{Bajardi20}. The application of $f(Q)$ gravity to the spherical symmetric configuration has also been explored and its effects can be demonstrated by the external and internal solution of compact stars \cite{Lin21}. Khyllep et al. \cite{Khyllep21} have performed the singularity analysis and dynamical system analysis of the model obtained in $f(Q)$ gravity and have shown the cosmological viability. Frusciante \cite{Frusciante21} has studied cosmological observables in the context of $f(Q)$ gravity. Anagnostopoulos et al. \cite{Anagnostopoulos21} have given the first evidence that the non-metricity $f(Q)$ gravity could challenge $\Lambda$CDM behaviour of the universe. To note here an extension of the non-metricity gravity is the $f(Q,T)$ gravity \cite{Xu19} that has been instrumental in providing late time cosmic acceleration models of the Universe \cite{Pati21a,Agrawal21,Pati21b}.\\

The motivation behind this work is to address the late time cosmic acceleration issue in some functional forms of $f(Q)$, described in the action of symmetric teleparallel gravity and to analyse the dynamical stability of the models. Mostly we focus on the mathematical simplification of $f(Q)$ gravity with the geometry modification. The paper is organised as follows: in Sec.\ref{sec:II}, we present a brief review of $f(Q)$ gravity along with the field equations in an FLRW space-time. In Sec.\ref{sec:III}, we have presented two cosmological models with two functional forms of $f(Q)$ and derived the dynamical parameters in an assumed form of the Hubble parameter. The  behaviour of energy conditions are also analysed for the respective model. In Sec.\ref{sec:IV}, the stability of the models are performed in the dynamical system approach whereas in Sec.\ref{sec:V}, the linear homogeneous perturbation approach has been adopted. The conclusions are given in Sec. \ref{sec:VI}.

\section{$f(Q)$ Gravity and Field Equations}\label{sec:II}

The metric tensor $g_{\mu \nu}$ is the generalization of gravitational potential whereas $\Gamma^{\lambda}_{\mu\nu}$ defines the covariant derivative and parallel transport. According to \cite{Hehl95,Ortin15}, the affine connection may be decomposed into the following three independent components as,
\begin{equation}\label{eq:1}
\Gamma^{\lambda}_{~\mu\nu} = \{^{\lambda}_{~\mu \nu}\} + K^{\lambda}_{~\mu \nu} + L^{\lambda}_{~\mu\nu}.
\end{equation}

In Eq. \eqref{eq:1}, the Levi-Civita connection of the metric tensor $g_{\mu \nu}$ is,  $\{^{\lambda}_{~\mu\nu}\} \equiv \frac{1}{2} g^{\lambda \alpha}\left({\partial}_{\mu} g_{\alpha \nu}+{\partial}_{\nu} g_{\alpha \mu}-{\partial}_{\alpha} g_{\mu \nu}\right)$. The contortion, $K^{\lambda}_{~\mu \nu}\equiv \frac{1}{2}\left( T^{\lambda}_{~\mu \nu}+T^{~\lambda}_{\mu~~\nu}+T^{~\lambda}_{\nu~~\mu}\right)$, where the torsion, $T^{\lambda}_{\mu \nu}\equiv \Gamma ^{\lambda}_{\mu\nu}- \Gamma ^{\lambda}_{\nu\mu}$. The disformation can be defined as, 
\begin{equation}\label{eq:2}
L^{\lambda}_{~\mu \nu} \equiv \frac{1}{2}(Q^{\lambda}_{~\mu \nu} - Q^{~\lambda}_{\mu~~\nu} - Q^{~\lambda}_{\nu~~\mu}),
\end{equation}
where the non-metricity, $Q_{\beta \mu\nu}\equiv \nabla_{\beta}g_{\mu \nu}$.

As mentioned in the introduction, the symmetric teleparallel gravity is free from the curvature and torsion and is defined by the non-metricity, we are intending to study the cosmological aspects of this non-metricity gravity. The action of $f(Q)$ gravity can be defined as,
\begin{equation}\label{eq:3}
S = \int \frac{1}{2}f(Q)\sqrt{-g}~d^{4}x + \int \mathcal{L}_{m}\sqrt{-g}~d^{4}x,
\end{equation}
where $g$ be the determinant of the metric $g_{\mu \nu}$ and $\mathcal{L}_{m}$, the matter Lagrangian. The non-metricity tensor is,
\begin{equation}\label{eq:4}
Q_{\beta \mu \nu} = \nabla_{\beta}g_{\mu \nu} = \frac{\partial g_{\mu \nu}}{\partial x^{\beta}},
\end{equation}
and its two traces as follows:
\begin{equation}\label{eq:5}
Q_{\beta} = Q^{~~\mu}_{\beta~~\mu}~~~~and~~~~\tilde{Q}^{\beta} = Q^{~\beta \mu}_{\mu}.
\end{equation}
The superpotential of the model is,
\begin{equation}\label{eq:6}
P^{\beta}_{~\mu \nu} \equiv -\frac{1}{4}Q^{\beta}_{~\mu \nu} + \frac{1}{4}\left(Q^{~\beta}_{\mu~\nu} + Q^{~\beta}_{\nu~~\mu}\right) + \frac{1}{4}Q^{\beta}g_{\mu \nu} - \frac{1}{8}\left(2 \tilde{Q}^{\beta}g_{\mu \nu} + {\delta^{\beta}_{\mu}Q_{\nu} + \delta^{\beta}_{\nu}Q_{\mu}} \right).
\end{equation}
The energy-momentum tensor is given by,
\begin{equation}\label{eq:7}
T_{\mu \nu} = -\frac{2}{\sqrt{-g}}\frac{\delta \sqrt{-g}\mathcal{L}_{m}}{\delta g^{\mu \nu}}.
\end{equation}
By varying the action \eqref{eq:3} with respect to the metric tensor $g_{\mu \nu}$, the field equations of $f(Q)$ gravity can be expressed as \cite{Jimenez18,Lazkoz19},
\begin{equation}\label{eq:8}
\frac{2}{\sqrt{-g}}\nabla_{\beta}(\sqrt{-g}f_{Q}P^{\beta}_{~\mu \nu}) + \frac{1}{2}g_{\mu \nu}f + f_{Q}(P_{\mu \beta \alpha}Q^{~~\beta \alpha}_{\nu} - 2Q_{\beta \alpha \mu}P^{\beta \alpha}_{~~\nu}) = -T_{\mu \nu},
\end{equation}
where $f_Q=\frac{\partial f}{\partial Q}$. 

To frame the cosmological model of the universe, we consider the spatially homogeneous, isotropic and flat FLRW  space-time, 
\begin{equation}\label{eq:9}
ds^2=-dt^2+a(t)^2(dx^2+dy^2+dz^2),
\end{equation}
and the energy momentum tensor as,
\begin{equation}\label{eq:10}
T_{\mu \nu} = (\rho + p)u_{\mu}u_{\nu} +pg_{\mu \nu}.
\end{equation}

The non-metricity, $Q=6H^2$ and the field equations of $f(Q)$ gravity in FLRW space-time can be obtained as \cite{Jimenez18},

\begin{eqnarray}
6f_QH^{2} - \frac{f}{2} &=& \rho,\label{eq:11}\\
(12f_{QQ}H^{2} + f_Q)\dot{H} &=& -\frac{1}{2}(\rho + p).\label{eq:12}
\end{eqnarray}
We take here, $8\pi G=c=1$. In case of standard matter, the continuity equation,   $\dot{\rho} = -3H(\rho + p)$ is satisfied and is also consistent with the cosmological equations. From Eqs. \eqref{eq:11}-\eqref{eq:12}, we can also establish the expression for the energy conditions in terms of the non-metricity function. For the energy momentum tensor, $T_{\mu \nu}=(\rho,-p,-p,-p)$, the general expressions for the energy conditions and their forms can be described as \cite{Novello08},
\begin{itemize}
\item Null Energy Condition [NEC]. For each null vector,

\begin{equation}
T_{ij}u^{i}u^{j} \geq 0 \Rightarrow \rho + p \geq 0 \Rightarrow  -2\left(12f_{QQ}H^2+f_Q\right)\dot{H}\geq 0.\label{eq:13}
\end{equation}

\item Weak Energy Condition [WEC]. For every time-like vector,
\begin{equation}
T_{ij}u^{i}u^{j} \geq 0 \Rightarrow \rho \geq 0~~and~~ \rho + p \geq 0 \Rightarrow -\frac{f}{2}+6f_QH^2\geq 0~~and~~-2\left(12f_{QQ}H^2+f_Q\right)\dot{H}\geq 0.  \label{eq:14}
\end{equation}

\item Strong Energy Condition (SEC). For any time-like vector,

\begin{equation}
\left(T_{ij}-\frac{1}{2}Tg_{ij}\right)u^{i}u^{j} \geq 0 \Rightarrow  \rho + 3p \geq 0\Rightarrow  f-6f_Q \left(\dot{H}+2H^2\right)-72f_{QQ}H^2\dot{H}\geq 0. \label{eq:15}
\end{equation}

\item Dominant Energy Condition (DEC). For any time like vector, 
\begin{equation}
T_{ij}u^{i}u^{j} \geq 0 \Rightarrow \rho - p \geq 0\Rightarrow-f+2f_Q(\dot{H}+6H^2)+24f_{QQ}H^2\dot{H}\geq 0~~and~~ T_{ij}u^{j}~\text{not~spacelike}. \label{eq:16}
\end{equation}

\end{itemize}

The energy conditions attributes to the fundamental casual and geodesic structure of space-time and can be useful in characterizing the attractive nature of gravity \cite{Capozziello19}. It has also a definite impact on the cosmic evolution, in particular the acceleration and deceleration, can be shaped by energy conditions. Both classical and quantum instabilities are triggered when these energy conditions are violated \cite{Carroll03}. In order to ensure physical validity, many cosmological models admitting $\omega\leq -1$ suffer from severe disabilities. This view discusses few ideas for avoiding instabilities \cite{Csaki05}.

Now, to find the pressure and energy density from the field equations \eqref{eq:11}-\eqref{eq:12}, the form of $f(Q)$ would be needed. In the following section, we have considered two cases pertaining to the functional forms $f(Q)$. 

\section{The Models and Dynamical Parameters}\label{sec:III}
In order to frame the cosmological model of the universe, it is usually preferred to obtain the exact solutions of the field equations. However, the highly non-linear equivalent Friedmann equations render the task some how difficult. In the present case, the equivalent Friedmann equations of the symmetric teleparallel gravity model are obtained to be highly non-linear and therefore, we consider some plausible assumptions to solve the system, first a form of $f(Q)$ and second a time varying deceleration parameter leading to a dynamically changing equation of state. Here, we assume two functional forms of $f(Q)$ and employ a specific type of the deceleration parameter of second degree as $q=-1+8n^{2}-12nt+3t^{2}$ \cite{Barky19}, with the Hubble parameters as, $H=\frac{1}{t(2n-t)(4n-t)}$. In this parametrization of the deceleration parameter, the geometrical behaviour of the model obtained in GR and explored the parameter $n=0.50$ in the range of the redshift $-1<z<4$ \cite{Barky19}. Also Nagpal and Pacif \cite{Nagpal21} have parametrized $q$ in $f(R,T)$ theory of gravity. However, with this parametric value of $n=0.50$, in the non-metricity gravity theory, the non-violation of strong energy condition compel us to choose some other parametric value. Therefore, we choose here $n=0.170,0.172,0.174$. Since the model decelerates or accelerates according to positive and negative behaviour of the deceleration parameter, we can retrieve here the accelerating behaviour for, $\frac{3t-\sqrt{2+3t^2}}{4}<n<\frac{3t-\sqrt{2+3t^2}}{4}$. Also the transition occurs at $t=2n\pm\sqrt{\frac{4n^2+1}{3}}$. The present value of the deceleration parameter for the representative values of the model parameter has been given in TABLE-I.\\

\begin{table}
\caption{Results for the deceleration parameter ($q_0$) and EoS parameter ($\omega_0$) at the present epoch}
\centering
\begin{tabular}{c|c|c|c}
\hline
\hline
$n$  & $q_0$   & $\omega_0$~~(Model I) & $\omega_0$~~(Model II) \\

\hline

$0.170$     &- 0.942 &- 0.961& - 0.923\\
$0.172$     &- 0.940 &- 0.960& - 0.921 \\
$0.174$	   &- 0.938 &- 0.959& - 0.919 \\
\hline
\end{tabular}
\end{table}

\subsection{Model I}
\begin{figure}[H]
\centering
\minipage{0.50\textwidth}
\includegraphics[width=\textwidth]{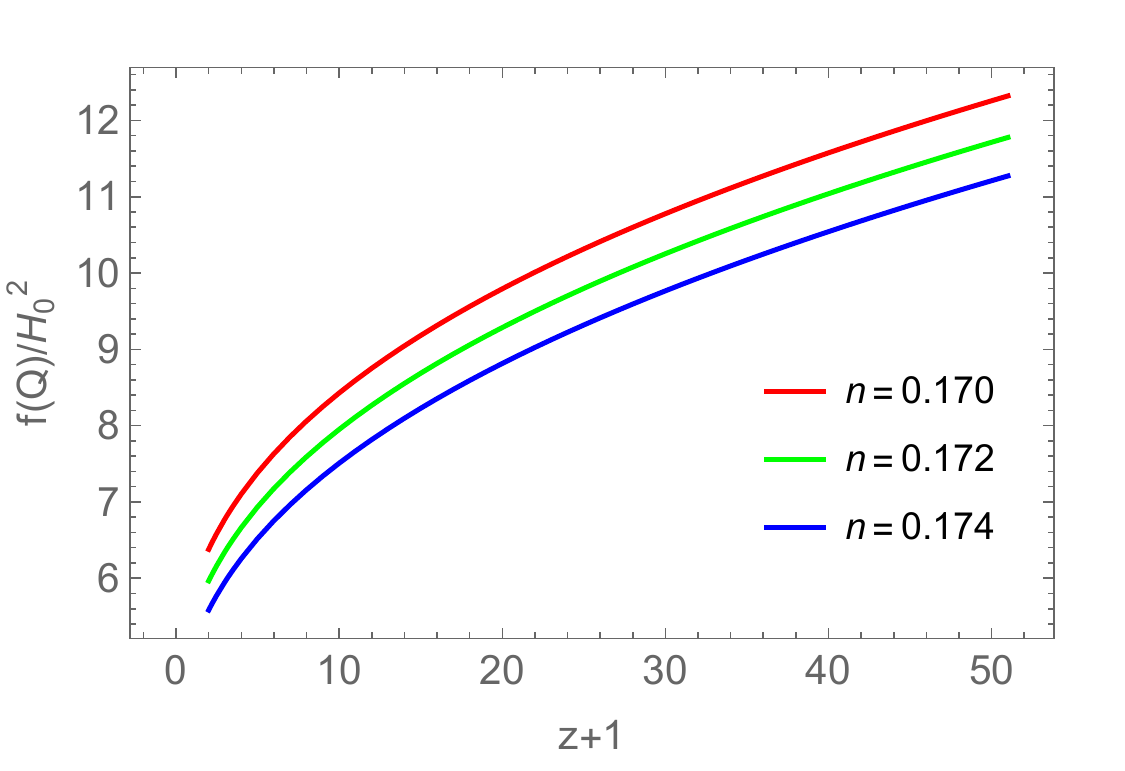}
\endminipage\hfill
\minipage{0.50\textwidth}
\includegraphics[width=\textwidth]{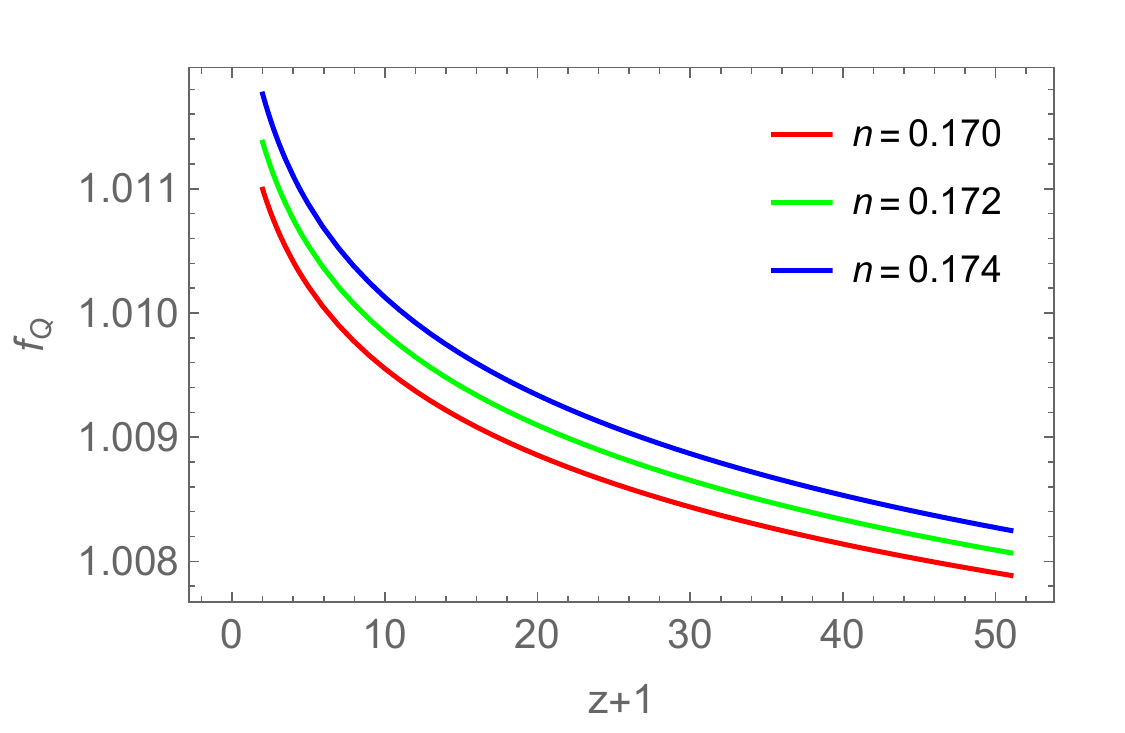}
\endminipage\hfill
\caption{Plot for the $\frac{f(Q)}{{H_0}^2}$ (Left Panel) and $f_Q$ (Right Panel) in redshift.}\label{FIG.2}
\end{figure}

As a first assumption of the non-metricity, we consider the specific form $f(Q)=Q+M\sqrt{Q}+C$, that represents the Friedmann equation in GR \cite{Jimenez20}. The constants $M$ and $C$ are respectively with dimensions of $mass$ and $mass^2$; the constant $C$ behaves as the cosmological constant whereas $M$ is a free parameter. Setting $C=\frac{M^2}{4}$, the form can be written as, $f(Q)=\left(\sqrt{Q}+\frac{M}{2}\right)^{2}$ \cite{Barros20}. We shall first assess the behaviour of the function $f(Q)$ with the assumed Hubble parameter. FIG. \ref{FIG.2} represents the plot for $\frac{f(Q)}{H_0^2}$ and $f_Q$ as function of  $z+1$, where $z$ is the redshift. Different values of the dimensionless parameter $M$, have been obtained earlier by analysing the evolution of the linear perturbation \cite{Frusciante21}, where  $H_0$ being the present time of the Hubble parameter. The Hubble parameter is related to the scale factor as $H=\frac
{\dot{a}}{a}$, so here we wish to analyse the $\frac{f(Q)}{H_0^2}$ and $f_Q$ as functions of redshift with different values of the scale factor parameter $n$. For $M=0$, it leads to the GR behaviour. However here we wish to see this behaviour with the change in the value of the scale factor parameter. The evolution of $\frac{f(Q)}{H_0^2}$ with the representative values of the parameter $n$ shows a decreasing behaviour and gradually vanishing at recent past. At the same time $f_Q$ starts from a small positive value and increases infinitely over time and remains entirely in the positive profile. This behaviour ensures an almost linear behaviour of the functional $f(Q)$.  

For the present choice of the non-metricity function, we obtain the pressure and energy density from Eqs. \eqref{eq:11} and \eqref{eq:12} as,

\begin{eqnarray}
p &=& \frac{M^{2}t^{2}(2n-t)^{2}(4n-t)^{2}+16(8n^{2}+3t^{2}-12nt)-24}{8 t^{2}(2n-t)^{2}(4n-t)^{2}},\label{eq:17}\\
\rho &=& \frac{24-M^{2}t^{2}(2n-t)^{2}(4n-t)^{2}}{8 t^{2}(2n-t)^{2}(4n-t)^{2}}.\label{eq:18}
\end{eqnarray}

It is well known that almost two-third of the energy budget of the universe is in the component with negative pressure and also known as the dark energy. In the literature, several models have been proposed to explain the effect of an exotic dark energy, the simplest and observationally consistent being the cosmological constant model. But due to fine tuning issue, the proposed dark energy models are mostly based on canonical, non-canonical scalar field and fluids \cite{Sami09,Silvestri09,Caldwell09} simulating a dynamical dark energy. So, to claim the late time acceleration, the value of equation of state (EoS) parameter, $\omega$ would be required, which is the ratio of pressure and energy density. The EoS parameter can be time dependent or it can be a constant also. Usually, the EoS parameter is obtained to be dynamically evolving for a rolling scalar field. Different cosmological observations in recent times constrained the present value of the EoS parameter as, $\omega=-1.29^{+0.15}_{-0.12}$ \cite{Valentino16}, $\omega=-1.3$ \cite{Vagnozzi20}, $\omega=-1.33^{+0.31}_{-0.42}$ \cite{Valentino21}. From Eqs. \eqref{eq:17}-\eqref{eq:18}, the EoS parameter may be obtained as, 

\begin{equation}
\omega = -1 +\frac{16(8n^{2}+3t^{2}-12nt)}{24-M^{2}t^{2}(2n-t)^{2}(4n-t)^{2}}. \label{eq:19}
\end{equation}

\begin{figure}[H]
\centering
\minipage{0.50\textwidth}
\includegraphics[width=\textwidth]{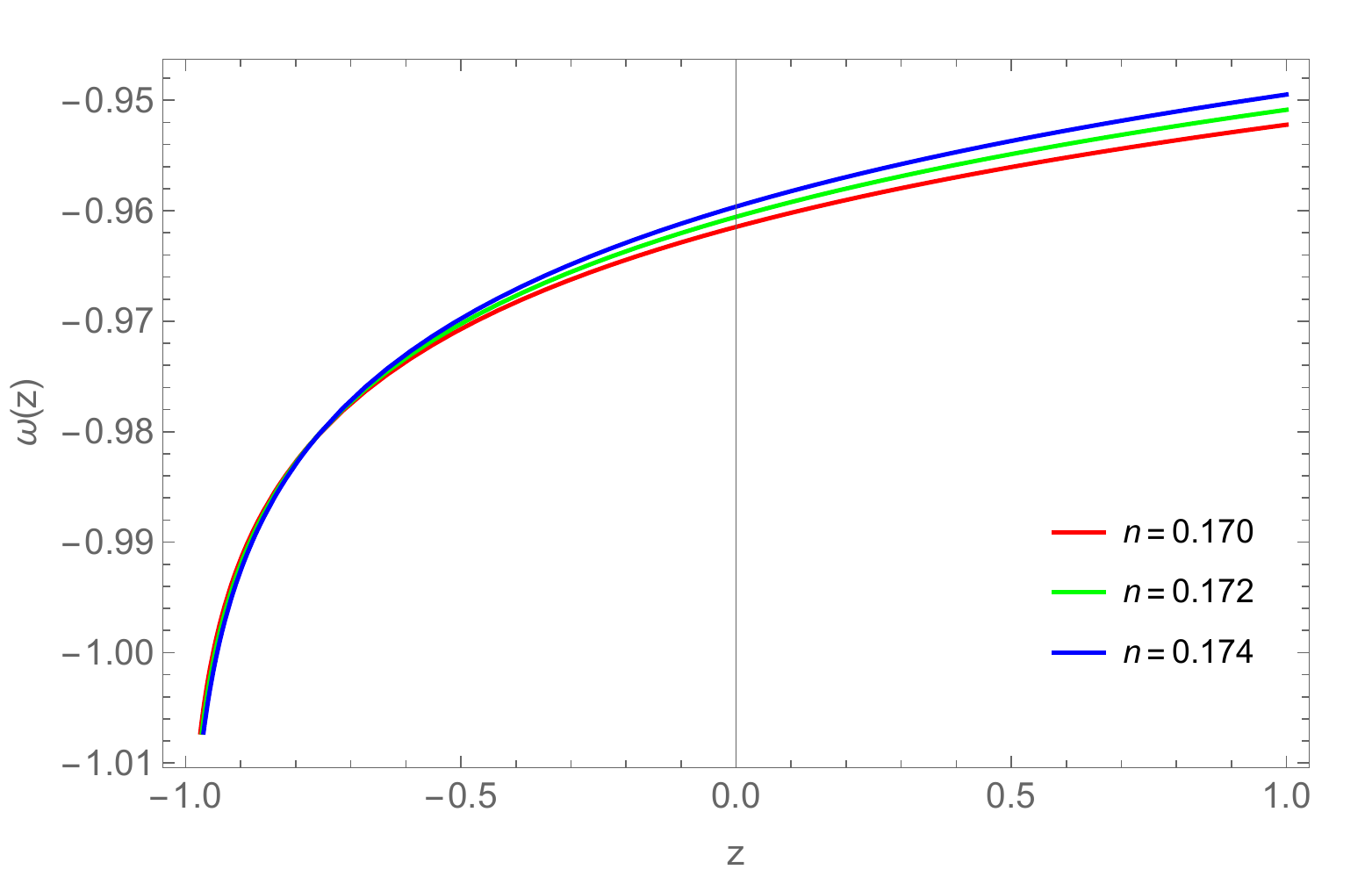}
\endminipage\hfill
\caption{Plot of EoS parameter in redshift. We have chosen, $M=4.0544$. }\label{FIG.3}
\end{figure}

We considered some representative values of the model parameters so as to keep the energy density positive throughout the evolution. Since no preferred range has been suggested by the cosmological observations, the chosen values are highly arbitrary in nature. For these choices of the model parameters, the energy density is observed to decrease gradually  maintaining the same behaviour in the immediate past and future time. The value of $M$ is chosen in such a way that the model can be supported by the cosmological observations data set \cite{Aghanim20}. Within the scope of the chosen models and parameters, the behaviour of EoS parameter is shown in FIG.\ref{FIG.3} which shows a decreasing behaviour from early epoch to future. At the present time, it crosses in the interval $[-0.961,-0.959]$. Subsequently it crosses the $\Lambda$CDM line at around $z\simeq -0.98$ and settled in the phantom phase [FIG.\ref{FIG.3}]. The predicted EoS parameter at the present epoch falls within the interval suggested by the cosmological observations.  The parameter $n$ affects the trajectory of the EoS parameter at the past epochs and pulls down the value of the EoS parameter with an increasing value of $n$. For, $n=0.170,0.172,0.174$, the present value of the  EoS parameter can be obtained as in TABLE I.  At the present time, it appears that, the model remains in a quintessence phase.

Energy conditions play important role in obtaining viable cosmological models. Here we analyse the energy conditions of the model using Eqs. \eqref{eq:13}-\eqref{eq:16} with the considered functional forms of $f(Q)$ and $H$. The energy conditions are obtained as,
\begin{eqnarray}
\rho + p &=& \frac{16(8n^{2}+3t^{2}-12nt)}{8t^{2}(2n-t)^{2}(4n-t)^{2}}, \nonumber \\
\rho + 3p &=& \frac{48(8n^{2}+3t^{2}-12nt)+2M^{2}t^{2}(2n-t)^{2}(4n-t)^{2}-48}{8t^{2}(2n-t)^{2}(4n-t)^{2}}, \nonumber \\
\rho - p &=& \frac{48-2M^{2}t^{2}(2n-t)^{2}(4n-t)^{2}-16(8n^{2}+3t^{2}-12nt)}{8 t^{2}(2n-t)^{2}(4n-t)^{2}}. \label{eq:20}
\end{eqnarray}
Since we are discussing here the viability of the functional form of $f(Q)$, in the context of modified theories of gravity the SEC should violate. At the same time, the DEC needs to be positive throughout and the behaviour of NEC depends on the present behaviour of the Universe. We have systematically investigated the energy conditions for the present model and observed that, while the SEC is violated throughout, the DEC is satisfied. Also, the NEC is found to be violated only in remote future.  

\subsection*{Model II}
\begin{figure}[H]
\centering
\minipage{0.50\textwidth}
\includegraphics[width=\textwidth]{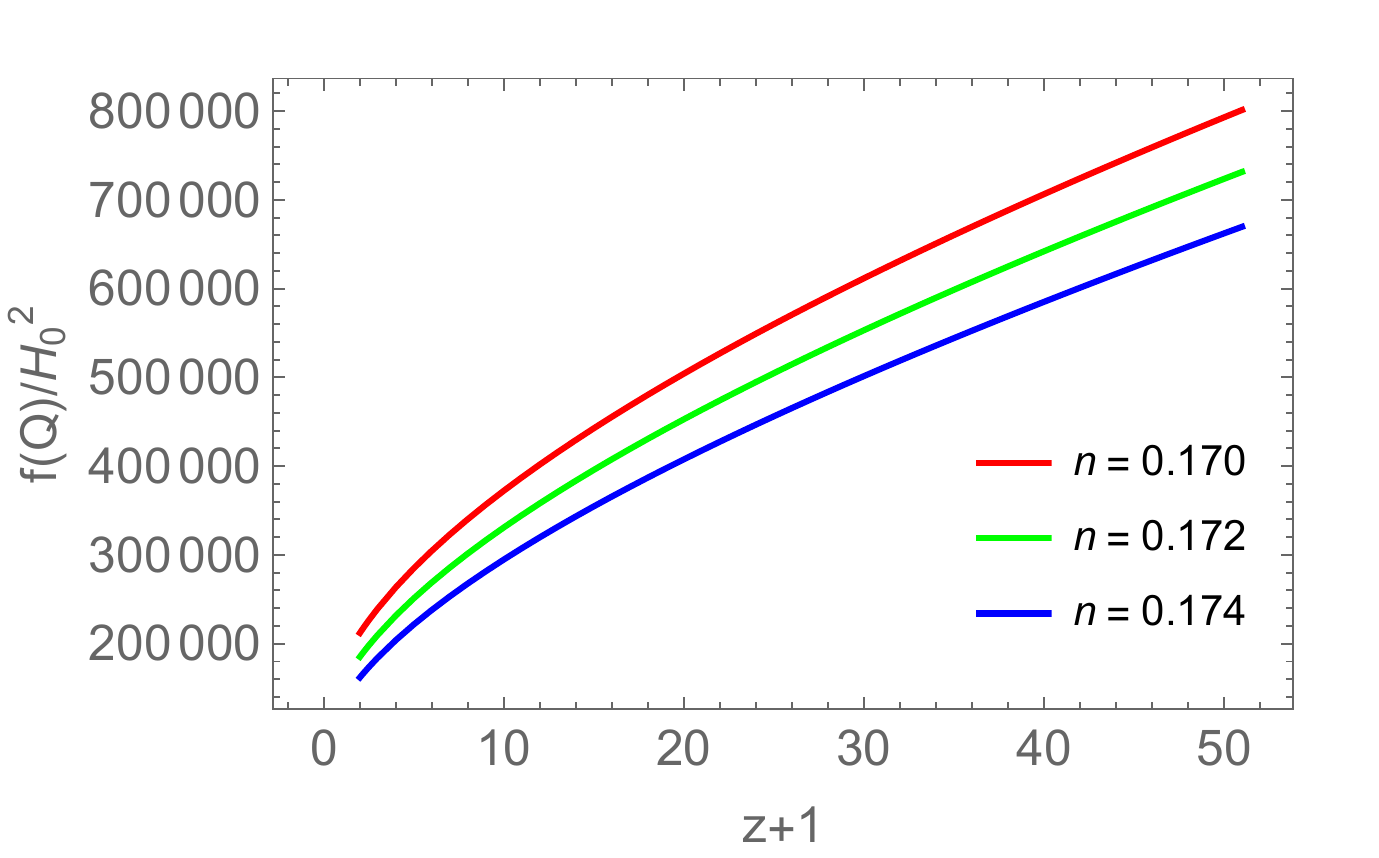}
\endminipage\hfill
\minipage{0.50\textwidth}
\includegraphics[width=\textwidth]{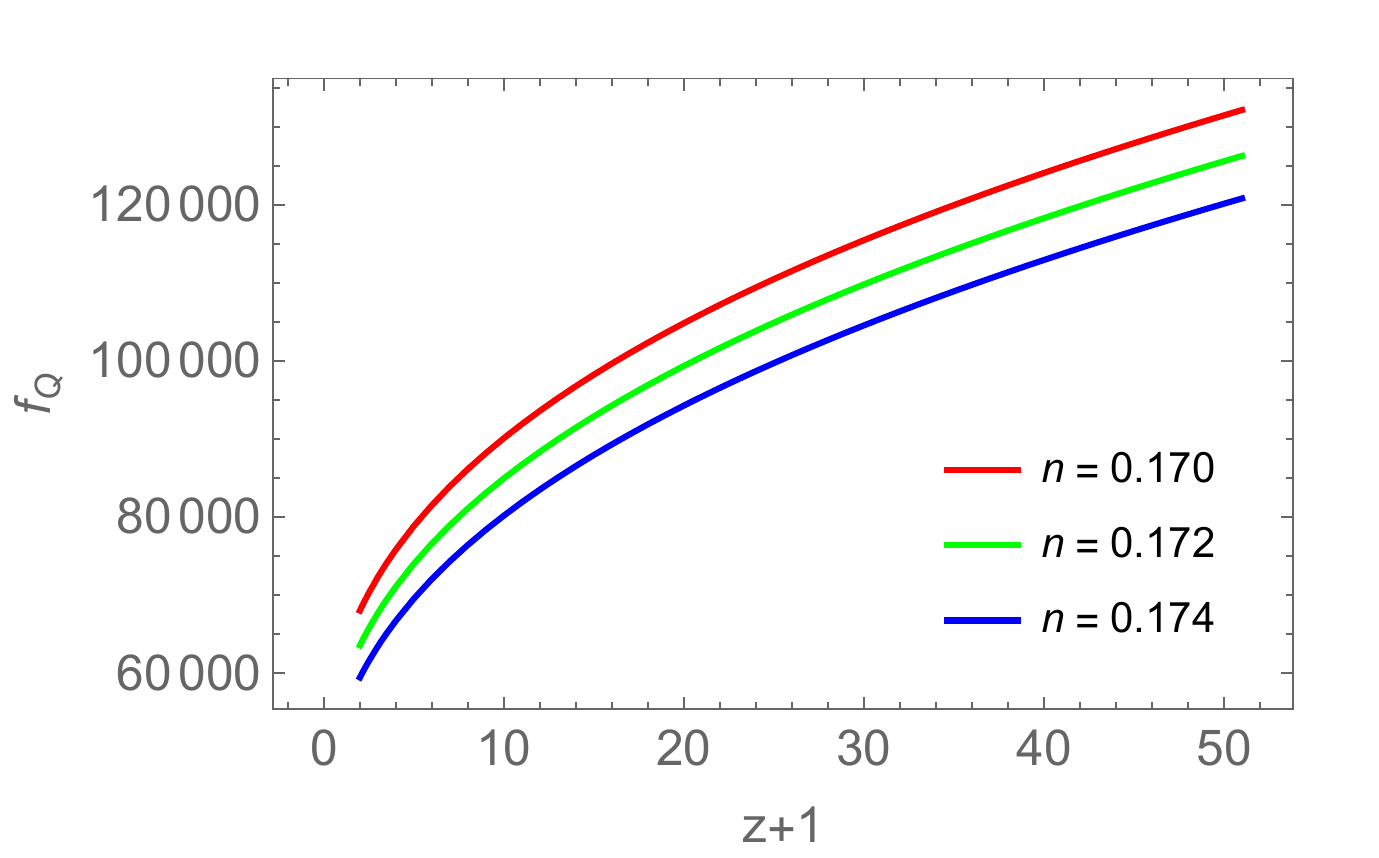}
\endminipage\hfill
\caption{Plots of $\frac{f(Q)}{{H_0}^2}$ (Left Panel) and $f_Q$ (Right Panel) in redshift. }\label{FIG.5}
\end{figure}

As a second example, we consider the non-metricity function as, $f(Q) = \left(Q + Q^{2}\right)$ \cite{Lu19}, which has a simple quadratic dependence on $Q$ and  we have $f_Q=1+2Q$. This heuristic example can be considered as analogous to Starobinsky model \cite{Starobinsky80} in the curvature based geometry with the model parameter as unity. We can see the behaviour of $\frac{f(Q)}{{H_0}^2}$ and $f_Q$ in FIG \ref{FIG.5}. In this model also, the normalized non-metricity function $\frac{f(Q)}{{H_0}^2}$  decreases gradually with cosmic time. However, the functional $f_Q$ becomes a decreasing function of the redshift. One may observe that, though at an initial epoch,  $f_Q$  differs for  different values of the parameter $n$, all the curves evolve to get merged very fast at late epoch. The motivation behind these plots is that the non-metricity scalar $Q$ at the FLRW background assumes $6H^2$ and to visualize and quantify the modifications. However, we can see that no major deviation is observed around the present time on different values of the parameter $n$. The derivative of $f(Q)$ observed to be the effective Planck mass and at large scale, it may impact the shape \cite{Jimenez20}.  

Now, substituting $f(Q)=Q+Q^2$, we can obtain the pressure, energy density from Eqns. \eqref{eq:11}-\eqref{eq:12} as,
\begin{eqnarray}
p &=&- \frac{2[36 + t^{2}(2n-t)^{2}(4n-t)^{2}](12nt-8n^{2}-3t^{2}) + 3 t^{2}(2n-t)^{2}(4n-t)^{2} + 54}{[t^{4}(2n-t)^{4}(4n-t)^{4}]}, \label{eq:21} \\
\rho &=& 3\left[\frac{18 + t^{2}(2n-t)^{2}(4n-t)^{2}}{t^{4}(2n-t)^{4}(4n-t)^{4}}\right].\label{eq:22}
\end{eqnarray}
and subsequently, the EoS parameter can be derived as, 
\begin{equation}\label{eq:23}
\omega = -1 -\frac{2}{3}\frac{[36 + t^{2}(2n-t)^{2}(4n-t)^{2}](12nt-8n^{2}-3t^{2})}{18 + t^{2}(2n-t)^{2}(4n-t)^{2}}.
\end{equation}
For the second model, the energy density and pressure are only dependent on the value of the parameter $n$, which controls the evolutionary aspects of the dynamical parameters. Hence we keep the same values of the parameter $n$ as in model I. In this process of fixing the parametric values, we obtained a positive energy density for the model.
\begin{figure}[H]
\centering
\minipage{0.50\textwidth}
\includegraphics[width=\textwidth]{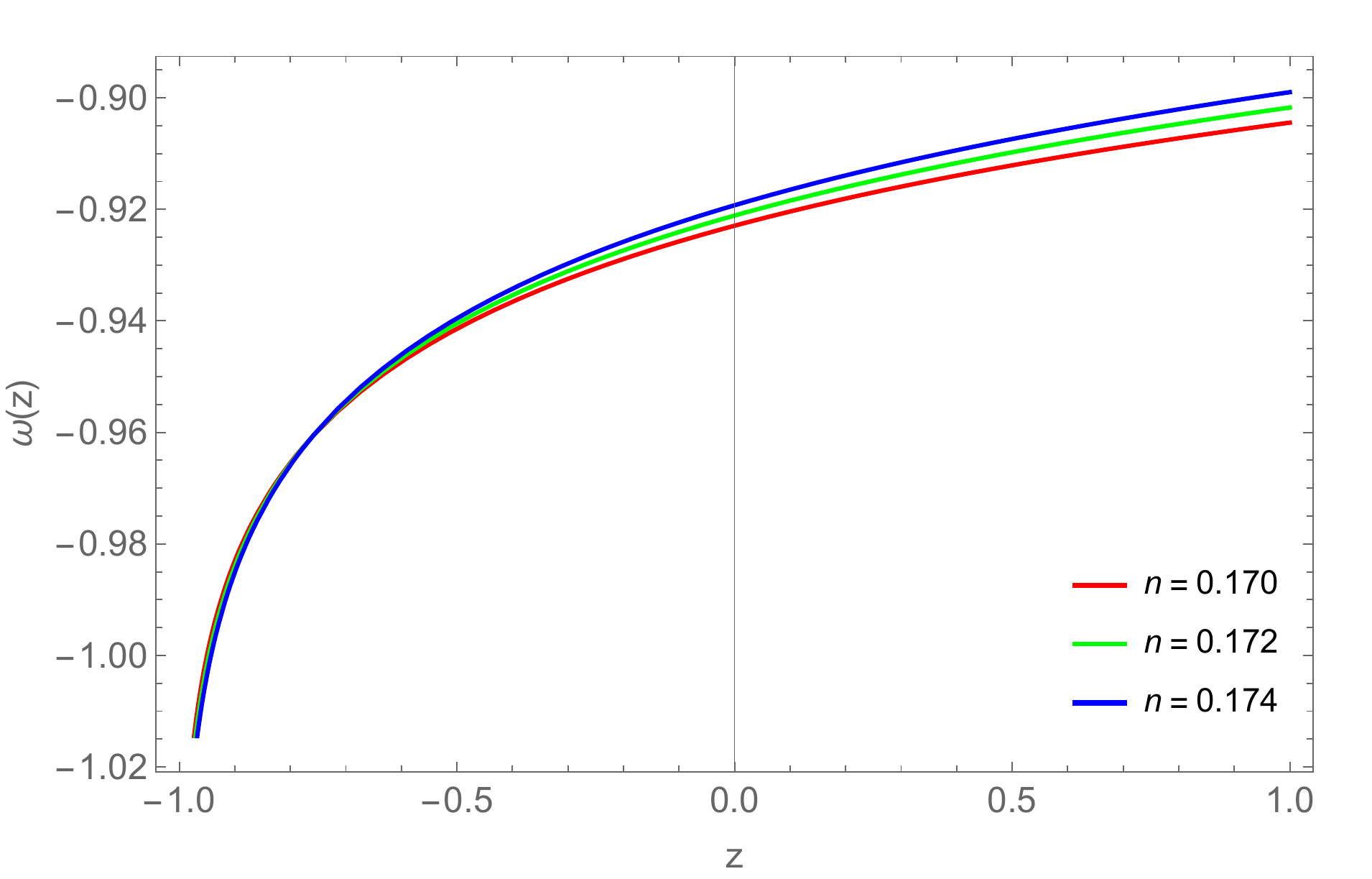}
\endminipage\hfill
\caption{Plot of EoS parameter in redshift. }\label{FIG.6}
\end{figure}

The EoS parameter decreases gradually as the redshift goes from positive to negative values [Fig. \ref{FIG.6}]. At the late phase all the curves remain in the phantom phase and  remain closer to the $\Lambda$CDM line. The predicted values of $\omega$ at $z=0$, have been listed in TABLE I which remain well within the observational limits. 

The energy conditions from \eqref{eq:13}-\eqref{eq:16} are obtained as,

\begin{eqnarray}
\rho+p &=& \frac{2(8n^{2}-12nt+3t^{2})[36+t^{2}(2n-t)^{2}(4n-t)^{2}]}{t^{4}(2n-t)^{4}(4n-t)^{4}}, \nonumber \\
\rho+3p &=& \frac{6(8n^{2}-12nt+3t^{2})[36+t^{2}(2n-t)^{2}(4n-t)^{2}]-6t^{2}(2n-t)^{2}(4n-t)^{2}-108}{t^{4}(2n-t)^{4}(4n-t)^{4}}, \nonumber \\
\rho-p &=& \frac{108+6t^{2}(2n-t)^{2}(4n-t)^{2}+2[36+t^{2}(2n-t)^{2}(4n-t)^{2}](12nt-8n^{2}-3t^{2})}{t^{4}(2n-t)^{4}(4n-t)^{4}}. \label{eq:24}
\end{eqnarray}

The energy conditions for the present model behave the same way as the previous model. While, the SEC is violated, the DEC is satisfied. However, the NEC is violated just after the present time. This further validates the problem in the context of extended theory of gravity.

\section{Dynamical System Approach in $f(Q)$ Gravity}\label{sec:IV}

In this section, we shall adhere to the dynamical system approach to analyse the stability of the systems, which can be performed by recasting the cosmological equations in the dynamical system \cite{Bahamonde18,Agostino18}. Here, we shall present the cosmological dynamical system in $f(Q)$ gravitational theory. If we take $f(Q)=Q + \Psi(Q)$ then Eqs. \eqref{eq:11} and \eqref{eq:12} becomes,
\begin{eqnarray}
3H^{2}-\frac{\Psi-2Q\Psi_{Q}}{2} &=& \rho, \label{eq:25}\\
(2Q\Psi_{QQ}+\Psi_{Q}+1)2\dot{H} + \frac{1}{2}(Q+2Q\Psi_{Q}-\Psi&)&=-p.\label{eq:26}
\end{eqnarray}
We consider that the universe is filled with dust and radiation fluids, so that,
\begin{equation}\label{eq:27}
\rho = \rho_{r} + \rho_{m}~~~~~~~p=\frac{1}{3}\rho_{r}
\end{equation}
where $\rho_{r}$ and $\rho_{m}$ are the energy densities of radiation and matter respectively. Also, we can express the Friedmann equations with the dark energy phase as,
\begin{eqnarray}
3H^{2} = \rho + \rho_{de},\label{eq:28}\\
2\dot{H} + 3H^{2} = -p-p_{de}.\label{eq:29}
\end{eqnarray}
where $\rho_{de}$ and $p_{de}$ respectively represents the dark energy density and pressure contribution caused by the geometry. So, geometrical dark energy ($\Psi(Q)$) density and pressure can be expressed as,
\begin{eqnarray}
\rho_{de} &=& \frac{\Psi}{2} - Q\Psi_{Q},\label{eq:30}\\
p_{de} &=& 2\dot{H}(2Q\Psi_{QQ} + \Psi_{Q}) + Q\Psi_{Q}-\frac{\Psi}{2}.\label{eq:31}
\end{eqnarray}
Now, we may introduce the density parameter pertaining to pressureless matter, radiation and dark energy respectively as, 

\begin{equation}\label{eq:32}
\Omega_{m} = \frac{\rho_{m}}{3H^{2}},~~~~~~~~~~~~~\Omega_{r} = \frac{\rho_{r}}{3H^{2}},~~~~~~~~~~~~~\Omega_{de} = \frac{\rho_{de}}{3H^{2}}.
\end{equation}
Subsequently, we obtain the effective EoS parameter and EoS parameter due to dark energy as,
\begin{eqnarray}
\omega_{eff}&=& -1+\frac{\Omega_{m} + \frac{4}{3}\Omega_{r}}{2Q\Psi_{QQ}+\Psi_{Q}+1},\label{eq:33}\\
\omega_{de} &=& -1 + \frac{4\dot{H}(2Q\Psi_{QQ}+\Psi_{Q})}{\Psi-2Q\Psi_{Q}}.\label{eq:34}
\end{eqnarray}
 
To analyse the dynamics of the cosmological models, we use the following dimensionless variables $x$ and $y$. Eqs. \eqref{eq:25}, \eqref{eq:26} and equation of continuity of motion are transformed into an autonomous system of first-order differential equations
\begin{equation}\label{eq:35}
x = \frac{\Psi-2Q\Psi_{Q}}{6H^{2}}, ~~~~~~~~~~~~~~~~~~~~~~~~~~~~ y = \frac{\rho_{r}}{3H^{2}}.
\end{equation}
If prime denotes differentiation with respect to the number of \textit{e}-folds of the Universe $ N = ln a$, then using chain rule, the equations of the model can be computed as,
\begin{equation}\label{eq:36}
\chi' = \frac{d\chi}{dN} = \frac{d\chi}{dt}.\frac{dt}{da}.\frac{da}{dN} = \frac{\dot{\chi}}{H}.
\end{equation}
Therefore, Eqs. \eqref{eq:25}, \eqref{eq:26} and equation of continuity can be transformed into the following dynamical system \cite{Khyllep21},
\begin{eqnarray}
x' &=& -2\frac{\dot{H}}{H^{2}}\left[\Psi_{Q}+2Q\Psi_{QQ} + x \right],\label{eq:37}\\
y' &=& -2y\left[2 + \frac{\dot{H}}{H^{2}}\right].\label{eq:38}
\end{eqnarray}
From Eqs.\eqref{eq:25} and \eqref{eq:28}, we obtain $ x + y = 1$, $\Omega_{r} = y$, $\Omega_{de} = x$ and $\Omega_{m} = 1-x-y $. Also we have,
\begin{equation}\label{eq:39}
\frac{\dot{H}}{H^{2}} = -\frac{1}{2}\frac{(3-3x+y)}{2Q\Psi_{QQ} + \Psi_{Q} + 1}.
\end{equation}
Next we shall analyse the models with the considered forms of $f(Q)$ in the problem.\\

For Model I, $f = (\sqrt{Q} + \frac{M}{2})^{2}$ such that $\Psi(Q) = M\sqrt{Q} + \frac{M^{2}}{4}$ and $2Q\Psi_{QQ} + \Psi_{Q} = 0 \Rightarrow \frac{\dot{H}}{H^{2}}= -\frac{1}{2}(3+y - 3x)$. Now, the dynamical system of the model can be obtained as,
\begin{eqnarray}
x' = x(3+y-3x),\label{eq:40}\\
y' = (y^{2}-3xy-y).\label{eq:41}
\end{eqnarray}
By extracting the critical points of the system through the solution of the equations $x' = 0$ and $y' = 0$, we will be able to analyse the cosmological dynamics of the model. The system contains three critical points as given in the Table-II below. Using the Jacobian matrix of a system at each critical point, we describe the stability nature of each critical point.
\begin{table}[H]
\caption{Critical points for the dynamical system corresponding to Model I.}
\centering
\begin{tabular}{c|c|c|c|c|c}
\hline
\hline
Point(x, y) &  $\Omega_{m}$ &  $\Omega_{r}$ &  $\Omega_{de}$ & $\omega_{eff}$ & Stability\\
\hline

A(0,0)     & 1 & 0 & 0 & 0 & saddle point\\

B(0,1)     & 0 & 1 & 0 & $\frac{1}{3}$ & unstable node\\

C(1,0)     & 0 & 0 & 1 & -1 & stable node\\
\hline
\end{tabular}
\end{table}
From the phase space analysis, it is clear that point $A(0,0)$ corresponds to matter dominated universe and $\omega_{eff} = 0$ shows that the universe is in decelerated phase. The unstable node $B(0,1)$ gives us decelerated, radiation dominated universe. Whereas $C(1,0)$ corresponds to accelerated universe. At $C(0,1)$, $\omega_{eff} = -1$. Moreover $C(0,1)$ is stable point showing dark energy dominated universe.
\begin{figure}[H]
\centering
\minipage{0.50\textwidth}
\includegraphics[width=\textwidth]{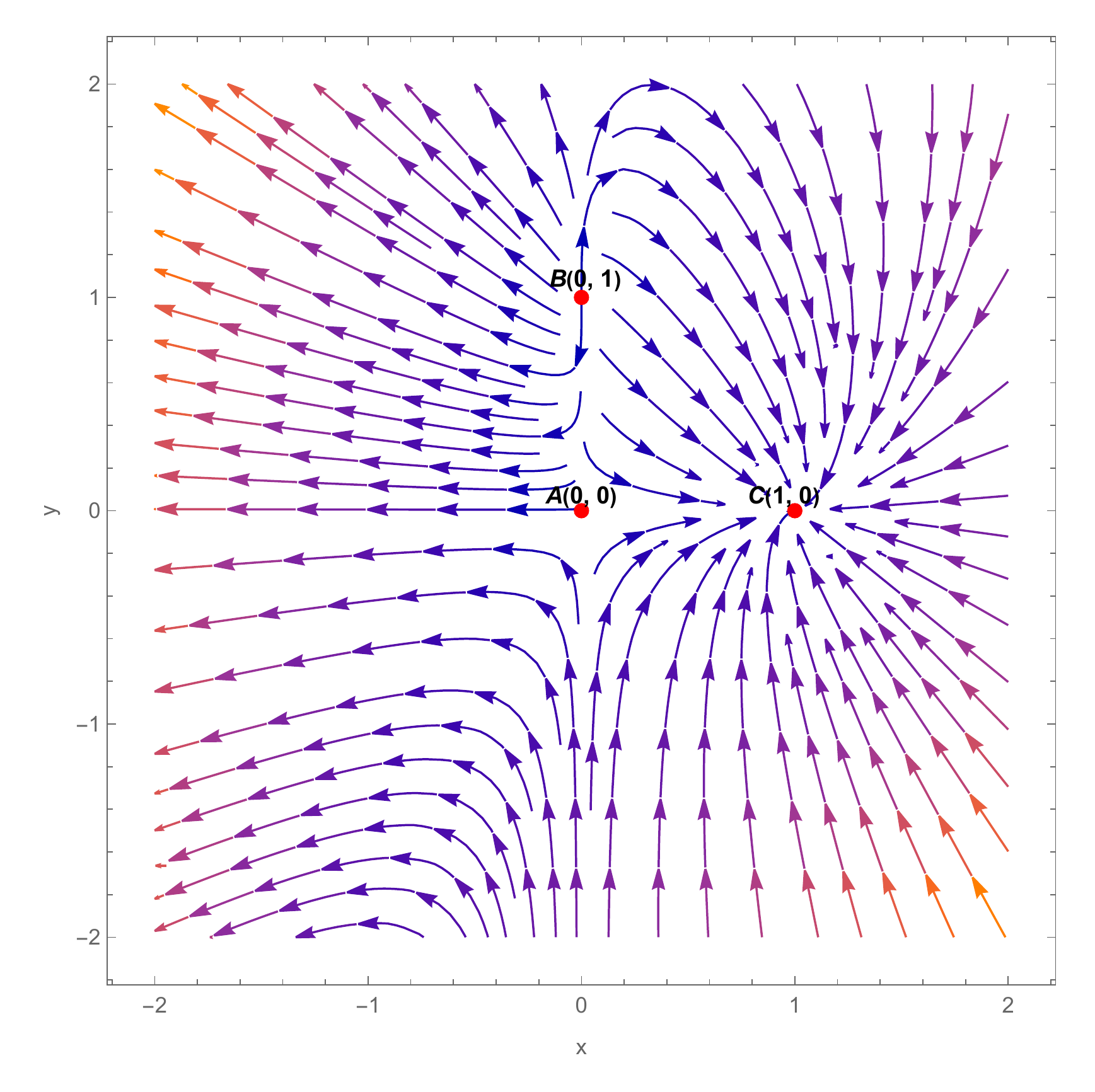}
\endminipage\hfill
\minipage{0.50\textwidth}
\includegraphics[width=\textwidth]{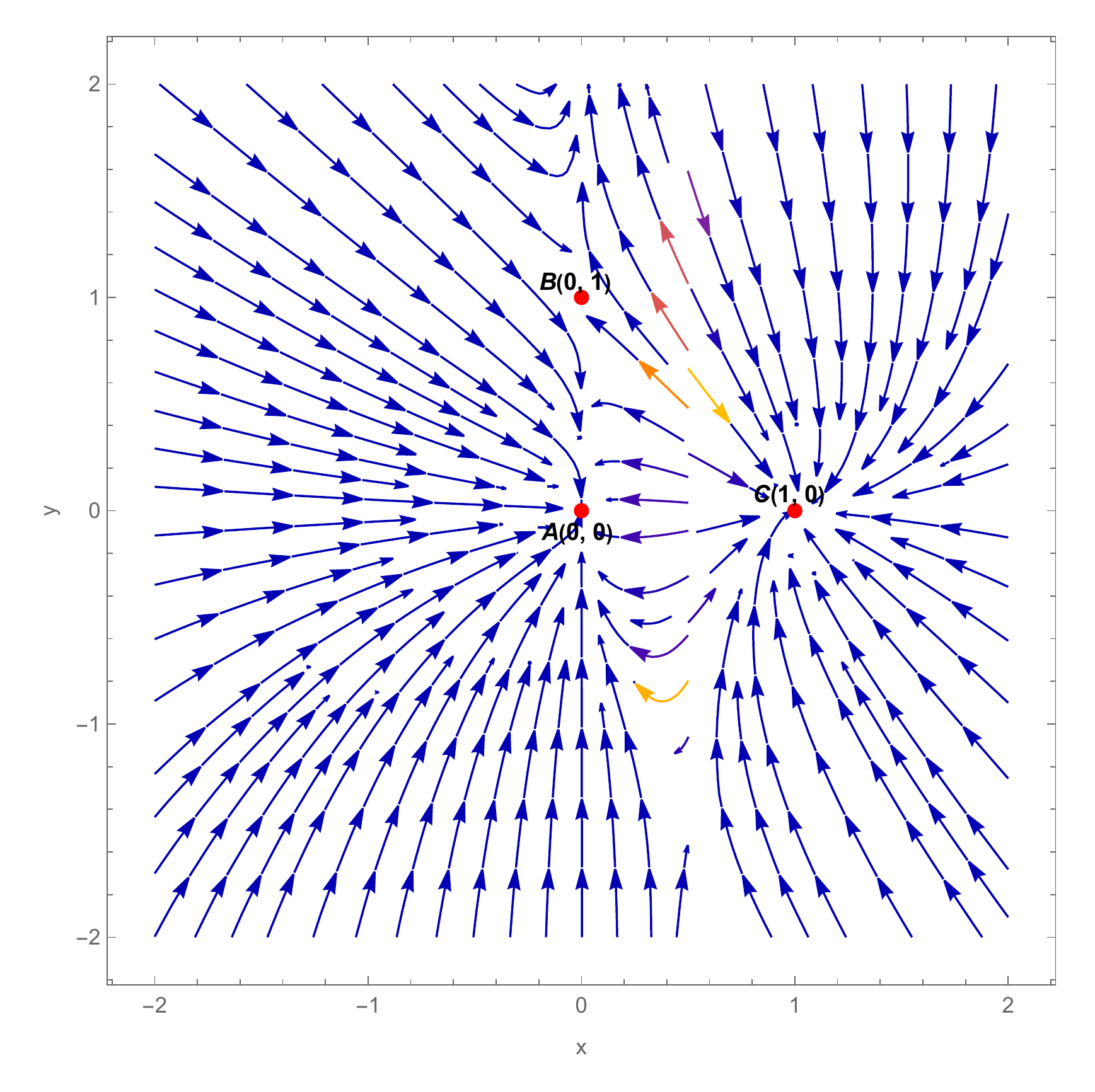}
\endminipage\hfill
\caption{Phase-space trajectories on the $x$-$y$ plane for $f(Q)$ gravity. The red dot corresponds to the critical point, Model I(left panel) and Model II(right panel).}  \label{FIG.8}
\end{figure}

Similarly, in Model II, $f(Q)=Q+Q^{2}$, hence $\Psi(Q) = Q^{2}$ and subsequently, $2Q\Psi_{QQ} + \Psi_{Q} = -2x \Rightarrow \frac{\dot{H}}{H^{2}}= \frac{1}{2}\frac{(3+y-3x)}{(2x-1)}$. We can derive the autonomous system for Model II as,
\begin{eqnarray}
x'&=& \frac{x(3+y-3x)}{(2x-1)},\label{eq:42}\\
y'&=& \frac{(y-5xy-y^{2})}{(2x-1)}.\label{eq:43}
\end{eqnarray}
Now, equating $x'=0$ and $y'=0$, the critical points can be obtained as in Table III,
\begin{table}[H]
\caption{Critical points for the dynamical system corresponding to Model II.}
\centering
\begin{tabular}{c|c|c|c|c|c}
\hline
\hline
Point(x, y) &  $\Omega_{m}$ &  $\Omega_{r}$ &  $\Omega_{de}$ & $\omega_{eff}$ & Stability\\
\hline

A(0,0)     & 1 & 0 & 0 & 0 & stable node\\

B(0,1)     & 0 & 1 & 0 & $\frac{1}{3}$ & saddle point\\

C(1,0)     & 0 & 0 & 1 & -1 & stable node\\
\hline
\end{tabular}
\end{table}
We, observe from Fig. \ref{FIG.8} that the dynamical system have two stable points: $A(0,0)$ and $C(1,0)$. Out of the two stable points, $A(0,0)$ corresponds to a matter dominated decelerated universe, whereas $C(1,0)$ suggests a dark energy dominated accelerated universe. The saddle point of dynamical system $B(0,1)$ shows a  radiation dominated decelerated universe. 

\section{ Stability Analysis Under Linear Homogeneous Perturbations} \label{sec:V}

In this section, we shall discuss the stability of the cosmological models against homogeneous and isotropic linear perturbations \cite{Cruz-Dombriz12,Farrugia16}. In the FLRW background, first order perturbations can be described as, 
\begin{equation}\label{eq:44}
H(t)\rightarrow H(t)(1+\delta),~~~~~~~~~~~~\rho(t)\rightarrow \rho(t)(1+\delta_{m}).
\end{equation}

Here, $H(t)$ and $\rho(t)$ represents the zero order quantities, and therefore satisfy Eqs. \eqref{eq:11}, \eqref{eq:12} and energy conservation equation. These quantities can also be denoted as $H_{b}$ and $\rho_{b}$ or $H_{0}$ and $\rho_{0}$ as observed in the literature. However, to distinguish them from the quantities evaluated at present time, these two notations are avoided here. Also, $\delta$ and $\delta_{m}$ respectively represent the isotropic deviation of the Hubble parameter and matter over-density. Now, the perturbation of the function $f$ and its derivatives can be obtained as,
\begin{equation}\label{eq:45}
\delta f = f_Q\delta Q,~~~~~~~~~~~~~ \delta f_Q = f_{QQ}\delta Q.
\end{equation}
Where $\delta f$ represents the first order perturbation of the variable $f$. Here, $\delta Q = 12H\delta H$. From energy conservation equation and from eq. \eqref{eq:11}, we get
\begin{eqnarray}
Q(2Qf_{QQ} + f_Q)\delta = \rho\delta_{m},\label{eq:46} \\
\dot{\delta}_{m} + 3H(1 + \omega)\delta = 0.\label{eq:47}
\end{eqnarray}
From \eqref{eq:46} and \eqref{eq:47} 
\begin{equation}\label{eq:48}
\dot{\delta}_{m} + \frac{3H(1 + \omega)\rho}{Q(2Qf_{QQ} + f_Q)}\delta_{m} = 0.
\end{equation}
Using eqs. \eqref{eq:12} in eqs.\eqref{eq:48}
\begin{eqnarray}
\dot{\delta}_{m} - \frac{\dot{H}}{H}\delta_{m} = 0,\label{eq:49}\\
\delta_{m} = cH. \label{eq:50}
\end{eqnarray}
$c$ is the constant of integration. Hence we found $\delta$ as
\begin{equation}\label{eq:51}
\delta = -k\frac{\dot{H}}{H}.
\end{equation}
\par Depending on the current value of $\delta_{m}$, the evolutions of both $\delta$ and  $\delta_{m}$ change. Stability is achieved as long as both $\delta$ and  $\delta_{m}$ decay with cosmic time. 

\begin{figure}[H]
\centering
\minipage{0.50\textwidth}
\includegraphics[width=7.5cm,height=5cm]{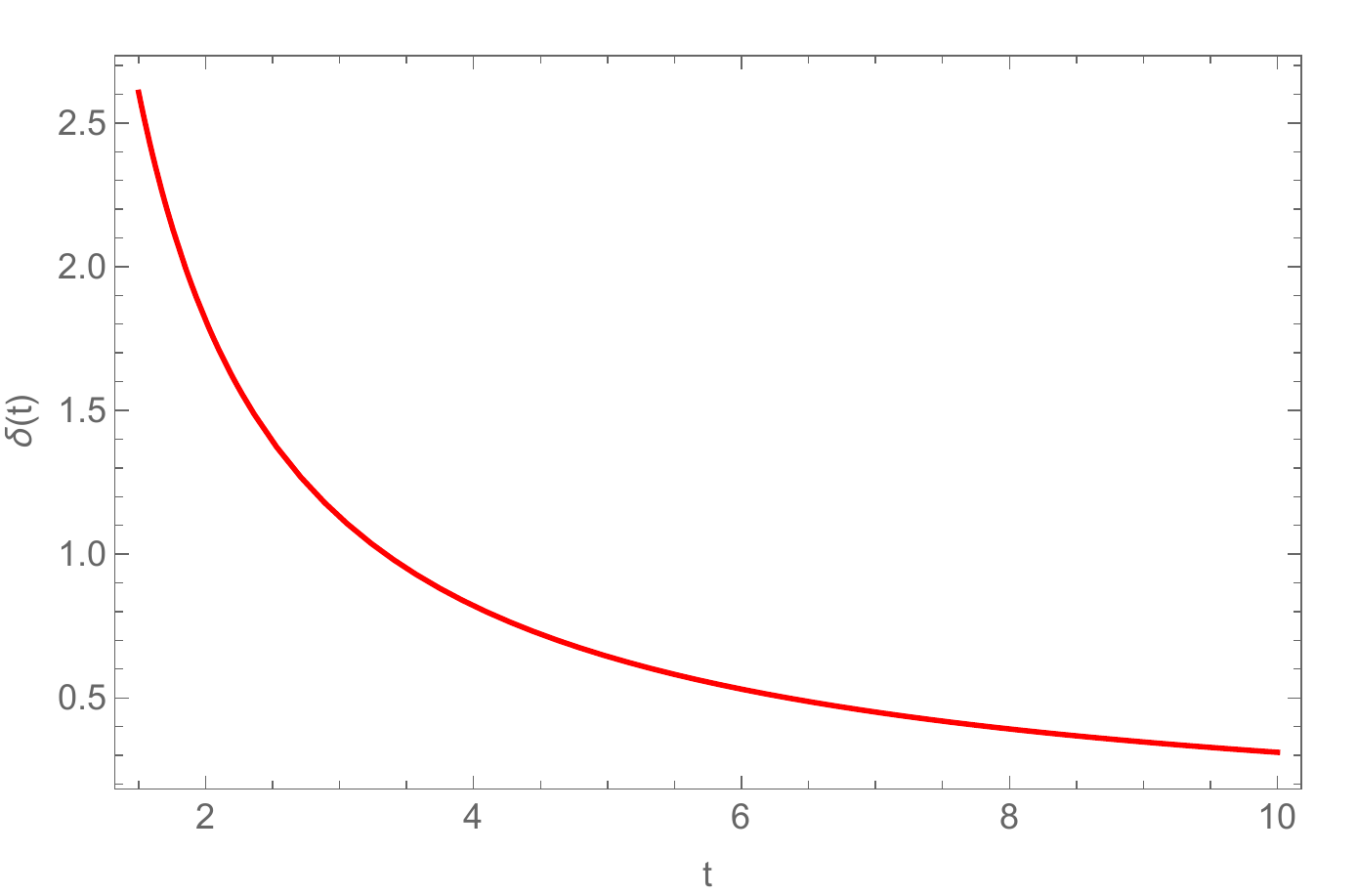}
\endminipage\hfill
\minipage{0.50\textwidth}
\includegraphics[width=7.5cm,height=5cm]{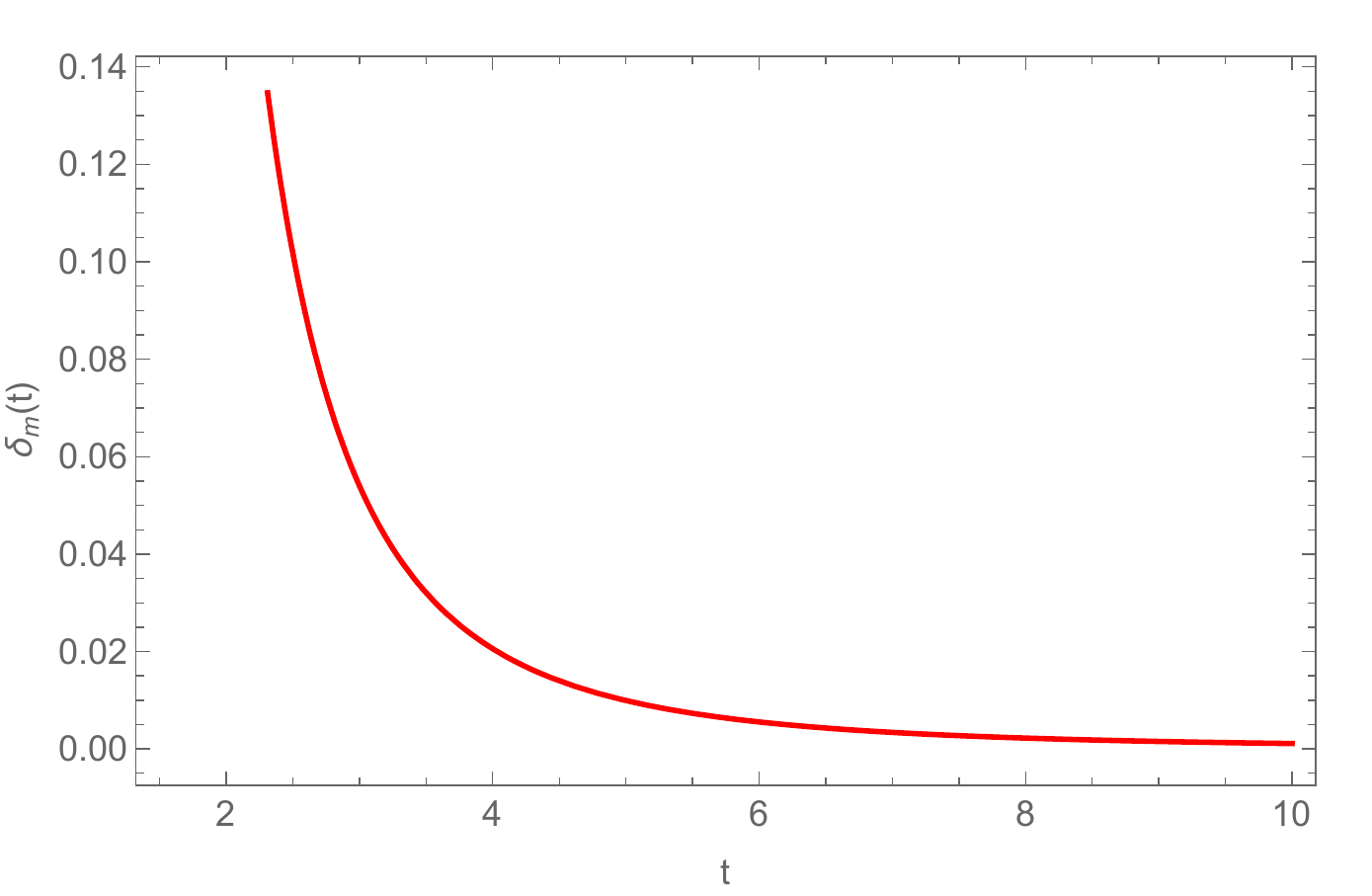}
\endminipage\hfill
\caption{Plot of $\delta$ (Left panel) and $\delta_{m}$ (Right panel) in cosmic time. The parameters, $n=0.170$, $k=1$, $c=1$}  \label{FIG.9}
\end{figure}

It can be observed from FIG. \ref{FIG.9} that both $\delta$ and $\delta_{m}$ are decaying over time and thereby confirming the stability of the model in the scalar perturbation approach also.

\section{Conclusion}\label{sec:VI}  

Several modified theories of gravity have been proposed in recent times to provide viable cosmological models that agrees with the late time cosmic acceleration issue. We have, in this paper, presented some cosmological models in a non-metricity based modified gravity. We have employed two forms of the function $f(Q)$ as $f(Q) = \left(\sqrt{Q} + \frac{M}{2}\right)^{2}$ and $f(Q) = Q + Q^{2}$. In Model I, the $\frac{f(Q)}{H_0^2}$ decreases and $f_Q$ increases gradually whereas in Model II,  both $\frac{f(Q)}{H_0^2}$ and and $f_Q$ decreases and vanishes at late time. Both the models are observed to show phantom behaviour at late times and the violation of strong energy conditions have been established. The numerical value of several cosmological observations have constrained the value of EoS parameter as: $\omega=-1.035^{+0.055}_{-0.059}$ (Supernovae Cosmology Project\cite{Amanullah10}), $\omega=-1.073^{+0.090}_{-0.089}$ (WMAP+CMB \cite{Hinshaw13}), $\omega=-1.03\pm0.03$ (Planck 2018 Results \cite{Aghanim20}). The model presented here exhibits the value of the EoS parameter (TABLE-I) which lies within these observational prescribed values. The present value of the deceleration parameter obtained through the cosmological observations as: $q_0=-1.08\pm0.29$ \cite{Camarena20} and the model provides the value within this range.

We have performed the phase space analysis by determining the critical points. In Model I, three critical points are obtained as $A(0,0)$, $B(0,1)$ and $C(1,0)$. At the critical point $C(1,0)$, the stability has been observed with the effective EoS parameter has value $-1$. We have encountered with one stable node, one unstable node and one saddle point. The stability of the model has been observed when the EoS parameter is in concordance with the $\Lambda$CDM behaviour. For the matter dominated phase it shows the saddle behaviour whereas in the radiation dominated phase, the instability occurs. The details are mentioned in Table II. For Model II, we wish to mention here that for $f(Q)=Q+\alpha Q^2$ \cite{Lu19} five critical points were determined and the stability was analysed. Here, we have performed the dynamical system approach in absence of the constant $\alpha$ to examine the stability behaviour. We have obtained three critical points and the stability of the model has been obtained at the point $C(1,0)$ that is in the dark energy era. The details are given in Table III.  The stability of the models are also studied through the discussion of the linear homogeneous perturbations of geometry and matter. The scalar perturbation approach, which is model dependent, has shown the stability of the model. Both the geometry and matter perturbations $\delta$ and $\delta_m$ decay over the time ensuring the stability of the model proposed.

In the conclusion, we wish to mention here that a symmetric teleparallel gravity, which has been formulated in non-metricity in place of curvature as in GR can be applied to provide viable models for late time acceleration in the universe. Theoretically the value of the EoS parameter has been significant to address the evolution history of the universe. In both the models, the behaviour of the dynamical parameters have been constrained by the scale factor and model  parameters. We hope the cosmological models based on non-metricity gravity may provide an alternative to the geometric dark energy models.

\section*{Acknowledgement}
LP acknowledges Department of Science and Technology (DST), Govt. of India, New Delhi for awarding INSPIRE fellowship (File No. DST/INSPIRE Fellowship/2019/IF190600) to
carry out the research work as Senior Research Fellow. BM and SKT acknowledge the support of IUCAA, Pune (India) through the visiting associateship program.

\end{document}